\documentclass[a4paper,fleqn,usenatbib]{mnras}
\usepackage{newtxtext,newtxmath}

\usepackage[T1]{fontenc}
\usepackage{ae,aecompl}

\usepackage{graphicx}	
\usepackage{amsmath}	
\usepackage{amssymb}    	
\usepackage{siunitx}
\usepackage{float}
\usepackage{xcolor}
\usepackage{enumitem}
\usepackage{multirow}

\newcommand{\gl}{\textsc{Gravlens}}
\newcommand{\lm}{\textsc{Lensmodel}}

\title{Testing the uniqueness of gravitational lens mass models}

\author[L. G. Walls \& L.L.R. Williams]{
Levi G. Walls,$^{1}$
Liliya L.R. Williams,$^{1}$\thanks{E-mail: llrw@astro.umn.edu}
\\
$^{1}$School of Physics and Astronomy, University of Minnesota, 116 Church Street, Minneapolis, MN 55455, USA\\
}

\date{Accepted XXX. Received YYY; in original form ZZZ}

\pubyear{2018}

\begin{document}
\label{firstpage}
\pagerange{\pageref{firstpage}--\pageref{lastpage}}
\maketitle

\begin{abstract}
The positions of images produced by the gravitational lensing of background sources provide unique insight in to galaxy-lens mass distribution. However, even quad images of extended sources are not able to fully characterize the central regions of the host galaxy. Most previous work has focused either on the radial density profile of the lenses or localized substructure clumps. Here, we concentrate on the azimuthal mass asymmetries near the image circle. The motivation for considering such mass inhomogeneities is that the transition between the central stellar dominated region and the outer dark matter dominated region, though well represented by a power law density profile, is unlikely to be featureless, and encodes information about the dynamical state and assembly history of galaxies. It also happens to roughly coincide with the Einstein radius. We ask if galaxies that have mass asymmetries beyond ellipticity can be modeled with simpler lenses, i.e., can complex mass distributions masquerade as simple elliptical+shear lenses?  Our preliminary study indicates that for galaxies with elliptical stellar and dark matter distributions, but with no mass asymmetry, and an extended source filling the diamond caustic, an elliptical+shear lens model can reproduce the images well, thereby hiding the potential complexity of the actual mass distribution. For galaxies with non-zero mass asymmetry, the answer depends on the size and brightness distribution of the source, and its location within the diamond caustic. In roughly half of the cases we considered the mass asymmetries can easily evade detection. 
\end{abstract}




\section{INTRODUCTION}

Multiple image, or strong gravitational lensing is a unique tool for determining the detailed mass distribution in galaxies. The very central few kpc of galaxies are dominated by baryons, in the form of stars. The density of stars falls off rapidly with distance from the center, and at larger radii the total mass becomes dominated by dark matter. 

The typical radius where the contributions of the two mass components are equal---what we will call the transition radius---is around 5-10 kpc, in projection. Despite the fact that the dominant mass component changes from stars to dark matter, strong lensing studies show that the 3D spherically averaged density profiles of galaxies are well represented by a single power law in radius, $\rho\propto r^{-\gamma}$, where $\gamma$ was estimated to be about $2$ \citep{koopmans09,barnabe11}. Because this transition is smooth, it is sometimes called the bulge-halo conspiracy \citep{vanalbada86}, in the sense that the stellar and the dark matter components conspire to keep the total mass density slope the same across the transition region. This smooth transition is also seen in numerical simulations \citep{schaller15a}, and is probably the consequence of the central regions of galaxies approaching relaxation. 

However, the transition region is not completely featureless. Both simulations \citep{schaller15a} and observations \citep{thomas07,chae14} show that with increasing distance from center, the total density slope becomes somewhat steeper, then shallower, and then steeper again. The presence of these features implies that full relaxation has not been achieved yet, even though the galaxies are likely in a stable mechanical equilibrium \citep{young16}. Both simulations \citep[e.g.,][]{nav04,springel08} and theory \citep{williams10} show that in a fully relaxed, collisionless system there should be no non-monotonic slope changes at these radii. 

If the central regions of galaxies are not fully relaxed, one should expect density perturbations, or asymmetries, not just in the radial, but also in the azimuthal directions. For example, if the ellipticity position angles of the dark matter and stellar distributions are misaligned, then the transition curve will not be elliptical. Most studies of the bulge-halo conspiracy have concentrated on the circularly (or spherically) averaged density profiles; in other words, used only the size of the Einstein radius in the analysis.  But lensing can also map out density variations in the projected azimuthal direction, because the positions of images in the polar angle around the lens center are sensitive to the azimuthal distribution of mass at the Einstein radius. 

By coincidence, multiple images of background sources form roughly around the transition radius \citep[see Figure 12 of][]{GomerWilliams2017}. Thus the structure of this region, both in the radial as well as azimuthal directions, is reflected in the image positions. Future surveys will deliver hundreds or thousands of quad lenses \citep{agnello15,finet12,oguri10}, allowing the study of redshift evolution of the transition region in a statistical sense, leading to a better understanding of galaxy relaxation.

We note that the mass asymmetries we are interested in are not the same as clumpy substructure, which has been detected through lens reconstruction, in several recent studies \citep{hez16,veg14,veg12,veg10}.

What is the best way to extract the information about the azimuthal mass distribution in a lens? One could carry out mass reconstructions of individual lenses, using either parametric, or free-form techniques.  The drawbacks of this approach are two-fold: (a) lensing degeneracies, for example the monopole degeneracy \citep{saha00,lie08,lie12}, can conceal the true features of the azimuthal mass distribution, especially when these features are higher order perturbations compared to pure ellipticity, and (b) individual lens modeling cannot take advantage of the fact that lenses belong to a population, and hence share some common properties. 

In this paper we examine the ability of mass modeling of individual lenses to extract information about the azimuthal mass distribution in the transition region. \cite{GomerWilliams2017} studied quad populations, with the same goal in mind.

It is known that most quad lenses are well modeled by elliptical+shear mass distributions. However, even the most sophisticated models, those that consist of a superposition of two different profiles---one representing dark matter, and the other, the distribution of stars \citep[e.g.][]{suyu09,gil17}---do not include the azimuthal complexities potentially associated with the transition region. Higher order perturbations in the galaxy isodensity contours may be present \citep{woldesenbet15,GomerWilliams2017}, but would elude mass modeling because of lensing degeneracies, which imply that there is an infinity of mass distributions that could reproduce the observed image properties. So an accurate fit to the images may not imply an accurately recovered mass distribution. 


The most well known is the mass sheet degeneracy \citep[MSD,][]{falco85}; its influence on the determination of $H_0$ from galaxy lenses was recently explored in \cite{xu16} and \cite{tagore18}. If one had a myriad of point sources at the same redshift, that covered the entire relevant portion of the source plane, and the multiple images of these sources were observed with infinite astrometric precision, then all degeneracies, except for MSD would be broken. These include the source plane transformation (SPT), which is a generalization of the MSD \citep{ss14}, and the monopole degeneracy, which reshapes the mass distribution between images. In addition, there could exist other degeneracies that have not been recognized and analyzed yet. These too would be broken by our idealized thought experiment.

Such an idealized experiment is not possible. The next best thing is to use extended sources with variable surface brightness distributions, such as galaxies, or host galaxies of QSOs; \cite{KKM01,BSK01}, but see \cite{saha00}. These provide useful, but far from complete coverage of the source plane. In addition to (i) limited spatial extent, other shortcomings of realistic observations are, (ii) finite size of detector pixels, translating in to imprecise astrometry, and (iii) faintness, resulting in low signal-to-noise ratio.  Compared to our idealised experiment, these allow degeneracies to creep in. In the present study, where we are interested in knowing how well, if at all, mass features of the transition region can be recovered through mass modeling, the most relevant degeneracy is not MSD or SPT, but probably the monopole degeneracy.

However, our goal is not to identify specific degeneracies, but ask if quads of extended sources, generated by mass distributions that have transition radius features, can be used to recover these features. Specifically, we ask a closely related question: can quads produced by complex galaxies be successfully modeled by simpler mass distributions? The answer will naturally depend on properties mentioned in (i), (ii), and (iii) above, which we explore in the paper. For some combinations of properties, the situation is promising, but in cases where simple mass distributions can indeed mimic more complex ones, the true features of the dark matter-stars transition region cannot be uncovered by mass modeling of individual lenses, and quad population studies, such as those presented in \citet{woldesenbet15} and \citet{GomerWilliams2017} are needed. 

\section{METHODOLOGY}	\label{sec:methods}

\subsection{Overview}

This section describes our analysis method in detail. We create galaxy mass models consisting of two components, dark matter and stars, with their centers offset in various directions, and by various amounts, up to $\sim15\%$ of the Einstein radius (or typically, $\sim 1h_{0.7}$~kpc), to generate non-elliptical mass distributions around the transition region. We do not claim that dark matter and stellar distributions are offset, but this is a convenient way to generate asymmetry at the image radius. Mass distribution very close to the lens centre, and hence the presence of the two mass peaks produced by the offset, are not relevant, because there are no lensing constraints there. \cite{GomerWilliams2017} showed that such asymmetry can reproduce the image polar angle properties of the population of observed galaxy quads, whereas the commonly used elliptical+shear models and $\Lambda$CDM substructure cannot. 

These are our 2-component ``observed'' galaxy lenses. A background cluster of point-sources, representing an extended source, is forward lensed, resulting in four islands of point-images. These are the ``observed'' extended images. The four images of the centre of the extended source---not the whole extended source---are then lens-modeled with a 1-component elliptical mass distribution centered on stars of the 2-component lens, and having an ``isothermal'' density slope, i.e., $\gamma=2$. We use the centre of the stellar distribution as the centre of the lens model, consistent with the common practice. Using just the central point-source is a simplification; one could use all the point-sources comprising the extended source to do the optimization, but that would be more cumbersome. This 1-component mass model has none of the complexities associated with the transition radius. Density slope is fixed at isothermal because that is what is commonly assumed in modeling. We then assess how well this 1-component lens can reproduce the extended source.

In all, we use four types of variables: parameters of the 2-component ``observed'' galaxy lens, stars-dark matter offset, source size, and source location within the diamond caustic. 

\subsection{Galaxy mass models}

Throughout this study we use the lensing software, \gl/\lm, made publicly available by Charles Keeton \citep{gravlens}. While \gl~and \lm~share the same underlying architecture, the former generates arbitrarily complex, parametric lensing galaxy (or galaxies) and lensed images, while the latter fits parametric mass models to lensed images. We used \gl~to generate 2-component galaxy lenses, and \lm~to model them using a 1-component galaxy.

From the \gl/\lm~catalogue of models \citep{KeetonMassModels,gravlens}, we chose to use boxy potentials ({\tt boxypot}) of the form 
\begin{equation}\label{eq:boxypot}
\phi = br^\alpha [1-\epsilon \cos 2(\theta - \theta_\epsilon)]^\beta
\end{equation}
because of their simplicity and computational efficiency. We used {\tt boxypots}\footnote{We note that \gl/\lm~software actually uses $\phi = br^\alpha [1-\epsilon \cos 2(\theta - \theta_\epsilon)]^{\alpha\beta}$ to do the calculations. We have corrected for this in the present work, and quote parameters as they appear in $\phi = br^\alpha [1-\epsilon \cos 2(\theta - \theta_\epsilon)]^\beta$, which is the form presented in the \cite{gravlens} manual.} to represent both the dark matter and the stellar distributions of the 2-component galaxies, as well as the 1-component fitted models. These parametric shapes have sufficient flexibility to be reasonable approximations to the density profiles of observed galaxies. 

To explore the parameter space, we investigate two 2-component models, called Galaxy 1 and Galaxy 2; their parameters are summarized in Table~\ref{tab:lens_comp}. The 1-component models that we use for lens fitting have fixed ``isothermal'' density slope, $\alpha=1$ ($\gamma=2$), and fixed $\beta=0.5$. Had we allowed these parameters to vary, better fits could have been obtained for the 1-component models. So we are being somewhat pessimistic about the ability of simple mass models to fit complex galaxies. Note that our 2-component galaxies do not have any external shear, $\gamma$, while 1-component models are allowed to have shear.

\begin{table}
	\centering
	\caption{Parameters for the 2-component galaxy lens potential; see eq.~\ref{eq:boxypot}. Adopting the same notation as \citet{gravlens}, $b$ is mass scaling, ${\epsilon}$ is ellipticity, and $\theta_{\epsilon}$ is the position angle of the ellipticity. All position angles are measured East of North. The density profile slope is $\alpha-2$. We do not include external shear in 2-component galaxies.}
	\label{tab:lens_comp}
	\begin{tabular}{ccccccc} 
		\hline
		galaxy & component & $b$  & $\epsilon$ & $\theta_{\epsilon} [^\circ]$ & $\beta$ & $\alpha$ \\
		\hline
		\multirow{2}{*}{Galaxy 1} & DM & 0.675 & 0.2 & -90 & 0.5 & 1.0\\
	                               	& Stars        & 4.05  & 0   & -90 & --  & 0.4\\
		\hline
		\multirow{2}{*}{Galaxy 2} & DM  & 0.5 & 0.2  & -70 & 0.5 & 1.1\\
	                        	& Stars        & 4.0 & 0.15 & -90 & 0.1 & 0.4\\
		\hline
	\end{tabular}
\end{table}

\begin{figure*}
	\includegraphics[width=\textwidth]{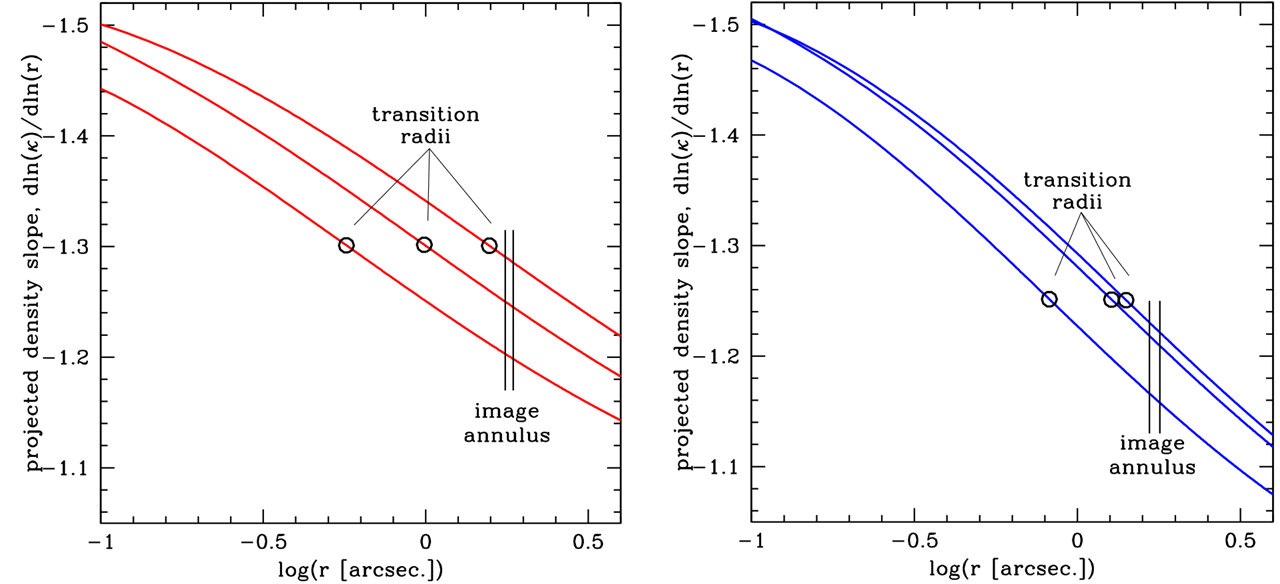}
	\caption{Double logarithmic density profile slopes for Galaxy 1 (left) and Galaxy 2 (right) 2-component lenses. The center of the stars is coincident with the center of the dark matter component. (We do not show profile slopes for the 2-component lenses where the dark matter and stars are offset; these would look somewhat different.) The three curves represent cuts along the $x-$axis, $y-$axis, and the $x=y$ diagonal line. The transition radius---where the projected densities of dark matter and stars are equal---along each one of these directions is shown with a black circle. The range of image radii is shown by the two vertical black lines.}
	\label{fig:slopes}
\end{figure*}

The projected double logarithmic slopes of the total---dark matter and stars---density profiles of the 2-component galaxies are shown in Fig.~\ref{fig:slopes}, for Galaxy 1 (left panel), and Galaxy 2 (right panel). Here, the centres of the dark matter and stellar distributions coincide. The cases with offset centres are not shown; these display a wider range of slope behaviours. The three curves represent cuts along the $x-$axis, $y-$axis, and the $x=y$ diagonal line. The transition radius---where the projected densities of dark matter and stars are equal---along each one of these directions is shown with a black circle. The range of image radii is shown by the two vertical black lines. 

The density slopes of 2-component galaxies are not constant, even though the slopes of dark matter and stellar distributions, taken individually, are constant. The central regions are dominated by stars, which have a steeper profile slope, while the outer regions are dominated by a shallower dark matter distribution. The typical inner slope is somewhat steeper than ``isothermal'', consistent with observations \citep{bell18,wong17,poci17}. 

The fact that density profiles are not power laws injects some complexity in to the galaxies' mass distribution. But the main property of the 2-component galaxies that makes them hard to model with a 1-component lens is the asymmetry at the image radius, resulting from the offset between the centres of the dark matter and stellar distributions. 


\subsection{Offsets: mimicking mass asymmetries at the transition radius}

\begin{figure*}
	\includegraphics[width=\textwidth]{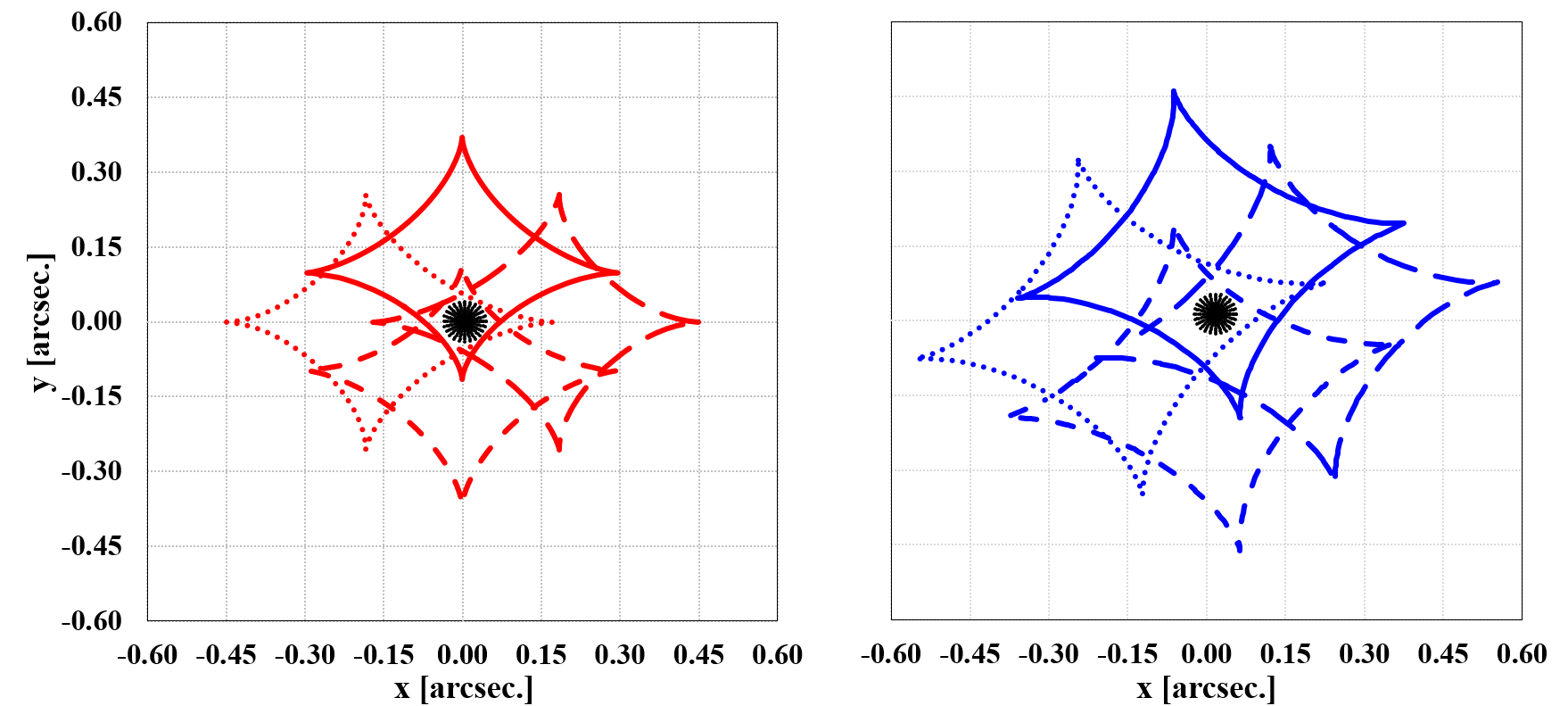}
	\caption{Caustics of the four 2-component lenses with the most offset stellar distributions. (The other 867 caustics are not shown.) The dark matter mass component is always centered at $(0,0)$. The centres of the stellar components of lenses whose caustics are shown with solid, long-dashed, short-dashed and dotted lines are at $(x,y) = (0\arcsec, 0.19\arcsec)$, $(0.26\arcsec, 0\arcsec)$, $(0\arcsec, -0.19\arcsec)$, and $(-0.26\arcsec, 0\arcsec)$, respectively. The black points are the cluster of point-sources that constitute the extended source, whose radius is $r_s=0.04\arcsec$, or $\sim 0.3h_{0.7}$~kpc for a typical source redshift of 3. The source size and the centres of the stellar mass component were chosen such that the whole of the extended source is guaranteed to produce quads for all 2-component lenses. {\it Left:} Galaxy 1 in Table ~\ref{tab:lens_comp}. Source centre is at $(x, y) = (0.0025\arcsec, 0.0025\arcsec)$. {\it Right:} Galaxy 2 in Table ~\ref{tab:lens_comp}. Source centre is at $(x, y) = (0.015\arcsec, 0.015\arcsec)$. }
	\label{fig:caustics}
\end{figure*}

The centre of the dark matter distribution is always at $(0\arcsec,~0\arcsec)$, so the centre of the stellar distribution is equivalent to the dark matter-stars offset. The centre of the extended source, shown as black dots in Fig.~\ref{fig:caustics}, is always at $(x, y) = (0.0025\arcsec, 0.0025\arcsec)$ for Galaxy 1 cases, and at $(x, y) = (0.015\arcsec, 0.015\arcsec)$ for Galaxy 2 cases. For Galaxy 1 and Galaxy 2 parameter sets separately, we create a suite of 2-component galaxies where the offset of the stellar distribution is varied in a 2D grid. The shape of the grid is not rectangular, but rather diamond shaped, as explained later.

We chose to focus on lenses that produce quads, as opposed to doubles, because quads provide more model constraining power than doubles. For a lens with a known centre, doubles supply 2 pieces of information ($x$ and $y$ of two images minus the 2 coordinates of the unobserved source), while quads supply 6. To guarantee the production of quads for all our lenses, we had to limit offsets. The four caustics that have the most extreme dark matter-stars offsets are shown in Fig.~\ref{fig:caustics}, for Galaxy 1 (left panel), and Galaxy 2 (right panel) parameter sets.  

As the centre of the stellar distribution is changed, the caustic shifts in the source plane, and its shape changes as well. Because the source position is fixed, it corresponds to different locations within the caustic for different offsets. Thus the offset and the relative source location are tied in our analysis: small offsets correspond to sources near the caustic center, while large offsets correspond to sources close to folds or cusps. Because of the symmetry, the former situation is easiest to model with a simple 1-component lens. Because of the complete lack of symmetry, the later situation is hardest to model. So even though we do not explore the full range of offset and source location combinations, our analysis includes the best, intermediate, and the worst cases, in terms of modeling difficulty.

\begin{figure}
	\includegraphics[width=\columnwidth]{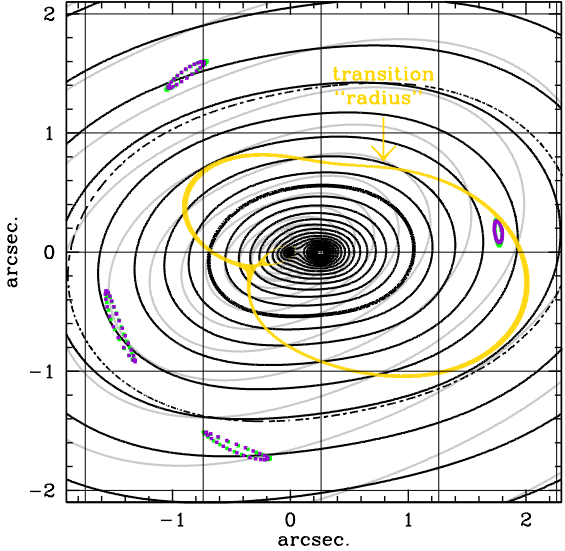}
        \vspace{-60pt}
	\caption{Isodensity contours of projected total (black) and dark matter (gray) density distributions for Galaxy 2, with the centre of the stellar distribution offset to ($0.26\arcsec,0\arcsec$). The dark matter is centred at ($0\arcsec,0\arcsec$), as in all cases in this paper. Contours are logarithmic, spaced by 0.1. The thick black and gray lines highlight the $\kappa=1$ contours. The grid and the dot-dashed black {\tt boxypot} ellipse are centred on the stars, the centre of the fitted 1-component lens model. The grid and the ellipse allow one to visually gauge the asymmetry of the 2-component galaxy. The four sets of 25 green points are the outer point-images of an extended source, located at $(0.015\arcsec, 0.015\arcsec)$, produced by the 2-component galaxy. The violet points are the corresponding images produced by the 1-component model, fitted to the central images (not shown) of that source. The unweighted (weighted) source-averaged $\rm {RMS}$ is $0.0099155\arcsec$ ($0.01299\arcsec$). The yellow band is the transition ``radius'', the region where the projected dark matter and stellar density differ by less than $1\%$, i.e. are nearly the same.}
	\label{fig:massgrid}
\end{figure}

\begin{figure*}
	\includegraphics[width=\textwidth]{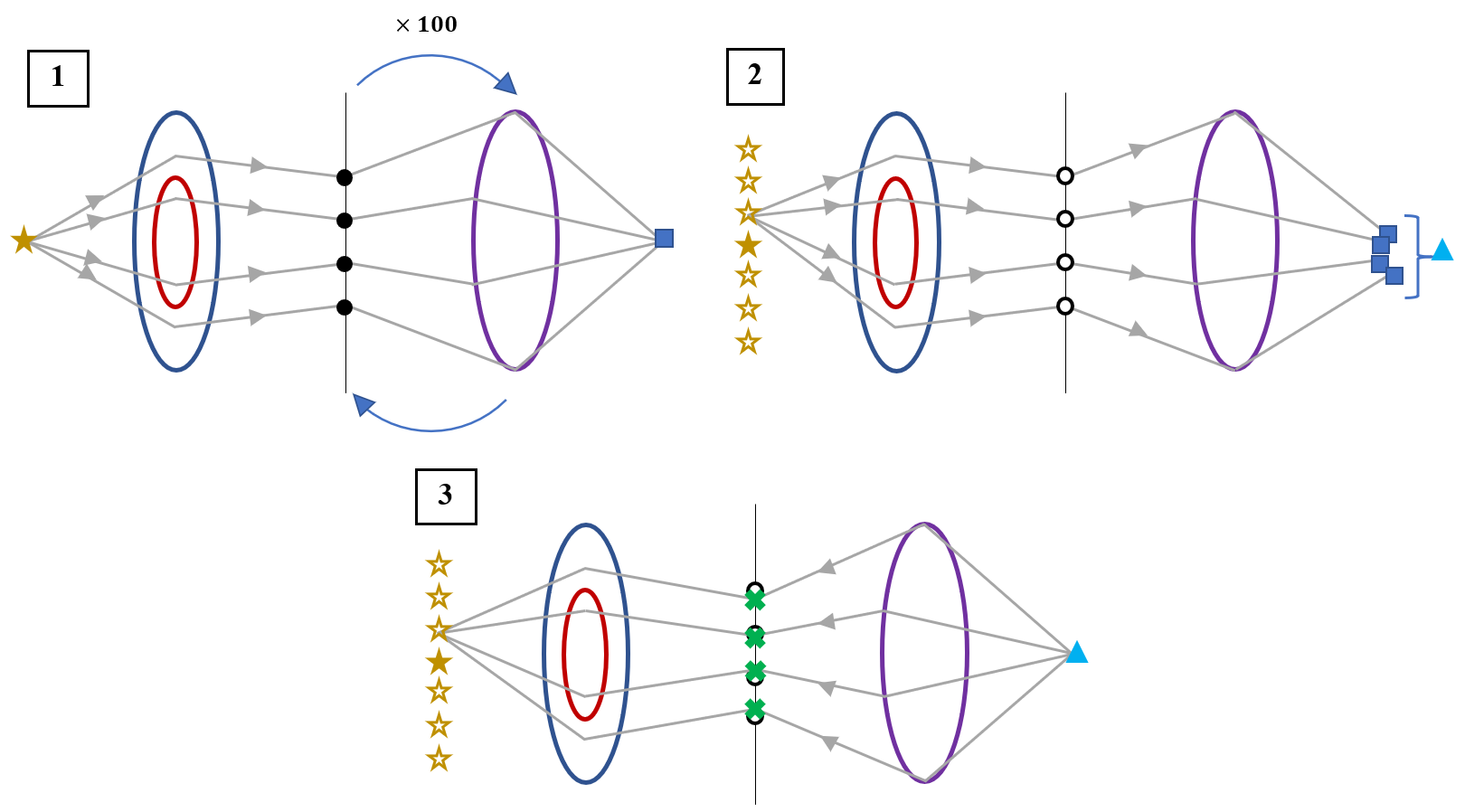}
	\caption{The schematic of our modeling procedure. {\framebox{\bf 1}}: A single point-source (filled, gold star) is forward-lensed through a 2-component system consisting of a dark matter halo (navy oval) and a stellar distribution (maroon oval). \lm~fits this quad with a 1-component model (purple oval), in 100 iterations. {\framebox{\bf 2}}: An extended source (empty, gold stars) is forward-lensed through the same 2-component system. Using the same  1-component lens (purple oval), the images (empty black dots) are back-projected and the average source position (cyan triangle) is calculated. {\framebox{\bf 3}}: The average source position is lensed to the image plane, where the $\rm {RMS}$ distance between the images from the 2-component galaxy (empty black dots) and the 1-component model (green crosses) is calculated. See Section~\ref{sec:model} for further discussion.}
	\label{fig:algorithm}
\end{figure*}

Fig.~\ref{fig:massgrid} shows an example of the 2-component galaxy mass distribution. The black lines are the total projected density contours of a Galaxy 2 lens, with the offset of the stellar distribution of ($0.26\arcsec,0\arcsec$). The gray lines show the dark matter component. The grid and the dot-dashed black {\tt boxypot} ellipse are centred on the stars, the centre of the fitted 1-component lens model. They allow one to visually gauge the asymmetry of the 2-component galaxy. The four sets of 25 green points are the outer point-images of an extended source, produced by the 2-component galaxy. The violet points are the corresponding images produced by the 1-component model. The yellow band is the transition ``radius'', the region where the ratio of the projected dark matter and stellar density differ by less than $1\%$. In this case, the transition region mass distribution deviates considerably from elliptical. In lenses with zero offset, the transition region band would look elliptical. 

\subsection{Modeling procedure}\label{sec:model}

We now describe the procedure we follow for a given position of the stellar distribution, i.e. a given offset. The offset is related to the amplitude of the transition region asymmetry. At the end of the three steps below, we quantify, in terms of ${\rm RMS}$, how well a 1-component lens reproduces an extended-source quad generated by a 2-component galaxy.  This procedure is repeated for each offset, which we vary in steps of $0.01 \arcsec$ in both spatial directions. Even for the most extreme offsets, the whole extended source is always quadruply imaged. This gives us a total 871 relative positions of the dark matter and stellar distributions.

The three steps below correspond to the three panels of Fig.~\ref{fig:algorithm}, a schematic of our modeling procedure.

\begin{enumerate}[leftmargin=*, listparindent=0.7cm, label={\framebox{{\bf\arabic*}}}]

\item {\bf Point source: forward lensing through a 2-component lens, and model fitting with a 1-component lens.} ~We use a 2-component lens, and a single point-source (filled gold star in all three panels of Fig.~\ref{fig:algorithm}) to generate 4 quad images (filled black dots) using \gl. These are the ``observed'' images of the centre of the extended source. Stars and dark matter mass components of the galaxy lens are represented schematically by the maroon and navy ovals, respectively, in all three panels. \lm~is then used to fit these 4 images with a 1-component {\tt boxypot}, centered on the stars. We ran \lm~for 100 iterations (blue arrows), optimizing over five parameters: the mass scaling ($b$), ellipticity ($\epsilon$), position angle of ellipticity ($\theta_{\epsilon}$), shear ($\gamma$), and position angle of shear ($\theta_\gamma$). One hundred iterations was a compromise between the robustness of the final model and computational time. The best fitting 1-component model is schematically shown as a purple oval in all 3 panels of in Fig.~\ref{fig:algorithm}. This same 1-component model is used in the following two steps.\\

\item {\bf Extended source: forward lensing through a 2-component lens, and tracing the images to the source plane through the 1-component lens.} ~Using the above point-source as the center, we then generated an extended source, represented by a large number of point-sources arranged in a spoke-like geometry (empty gold stars in Fig.~\ref{fig:algorithm}). The radius of the extended source is $r_s=0.04\arcsec$ ($\sim 0.3h_{0.7}$~kpc for a typical source redshift of 3), and its center is slightly misaligned with the centre of the dark matter of the 2-component lens. It contains 251 individual point-sources: 25 spokes (one every $15^\circ$), each with 10 point-sources, plus the central point-source. Using the 2-component lens, we forward lensed these sources to produce 251 quads, represented schematically by empty black dots. These are the ``observed'' extended images. These quads were then back-traced from the lens plane, through the 1-component lens obtained earlier, to the source plane. The four back-traced images of each quad (blue filled squares in Fig.~\ref{fig:algorithm}) were then averaged to produce the average source positions; one of these is schematically represented by a cyan triangle. Note that no lens fitting is done in this step.\\

\item {\bf Image comparison.} ~Each of the 251 averaged point-sources was forward lensed by the same 1-component lens, to generate a quad in the lens plane (green crosses). The image positions were then compared to the image positions of the corresponding ``observed'' quads. The comparison was done using the root mean square,
\begin{equation}	\label{eq:rms}
(\rm {RMS})^{2} = \frac{1}{4} \displaystyle\sum_{m=1}^{4} |\bmath{\theta}_{\rm obs,m} - \bmath{\theta}_{\rm mod,m}|^{2},
\end{equation}	
for each point-image quad of the extended source. 

\end{enumerate}


\begin{figure*}
	\includegraphics[width=\textwidth]{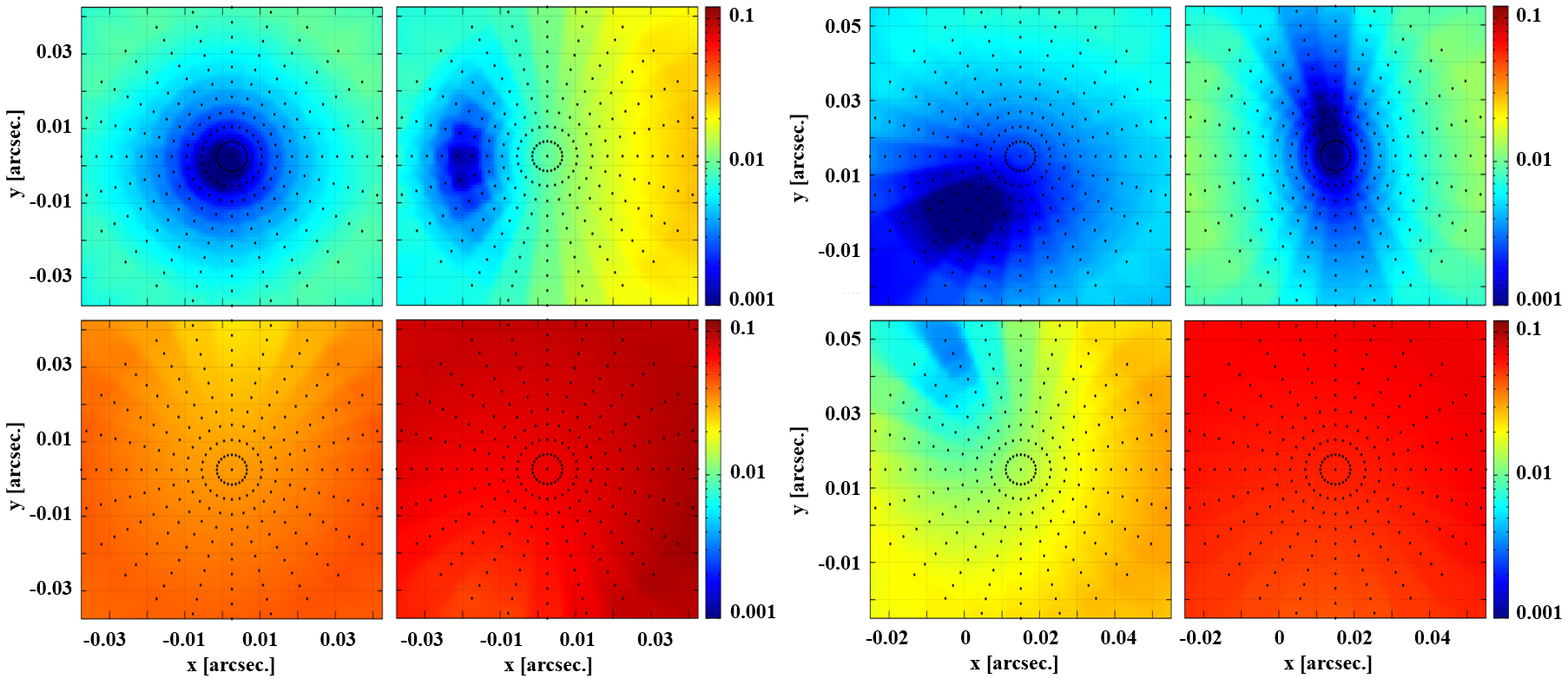}
    \caption{The color scale represents the $\rm {RMS}$ values for each of the point-sources (black dots) in the extended source centered at $(x,y)=(0.0025\arcsec, 0.0025\arcsec)$ for Galaxy 1 (leftmost group of four) and $(x,y)=(0.015\arcsec, 0.015\arcsec)$ for Galaxy 2 (rightmost group of four). The centres of the stellar distributions (i.e., offsets) are located at 
$(x,y)=(0\arcsec,0\arcsec)$, $(0.03\arcsec, 0\arcsec)$, $(0.04\arcsec,-0.04\arcsec)$, and $(-0.06\arcsec, 0.11\arcsec)$ for Galaxy 1 panels, and at  $(x,y)=(0\arcsec,0\arcsec)$, $(-0.04\arcsec,-0.01\arcsec)$, $(0.04\arcsec,-0.04\arcsec)$, and $(-0.01\arcsec, 0.15\arcsec)$ for Galaxy 2 panels. The data were interpolated between each of the point-sources. The $\rm {RMS}$ colour scale is logarithmic.}
    \label{fig:plot_src}
\end{figure*}

\section{RESULTS}	\label{sec:results}

In this section we look at how well a 1-component lens, fitted to the central point of an extended source, reproduces the ``observed'' quad images of that extended source. 

Our comparison is guided by the following consideration. If a point-image, which is a part of an extended ``observed'' image, is within one HST ACS pixel of the corresponding modeled point-image, then the difference between them will not be detected. If this is true, on average, for all the point-images of an extended source, then the 1-component lens reproduces the images of the 2-component ``observed'' galaxy well, and the true mass complexities of the 2-component lens have evaded modeling. In the next subsection we carry out the comparison on the point-image basis, using eq.~\ref{eq:rms}, as well as based on $\rm {RMS}$ averaged over the extended source.  

The ACS pixel is $0.05\arcsec$, and that of WFC-UVIS is $0.03\arcsec$, and can be effectively reduced to somewhat smaller values for drizzled images. The actual pixel size used can depend on the quality of the data. For example, \cite{ding17} used pixel size of $0.0325\arcsec$, or $0.0217\arcsec$, for their lenses, while \cite{tag18} assumed pixel size of $0.05\arcsec$ in their modelling work. We will use $0.03\arcsec$ as a threshold for deciding if the images are well fit by a model. 

\begin{figure*}
	\includegraphics[width=\textwidth]{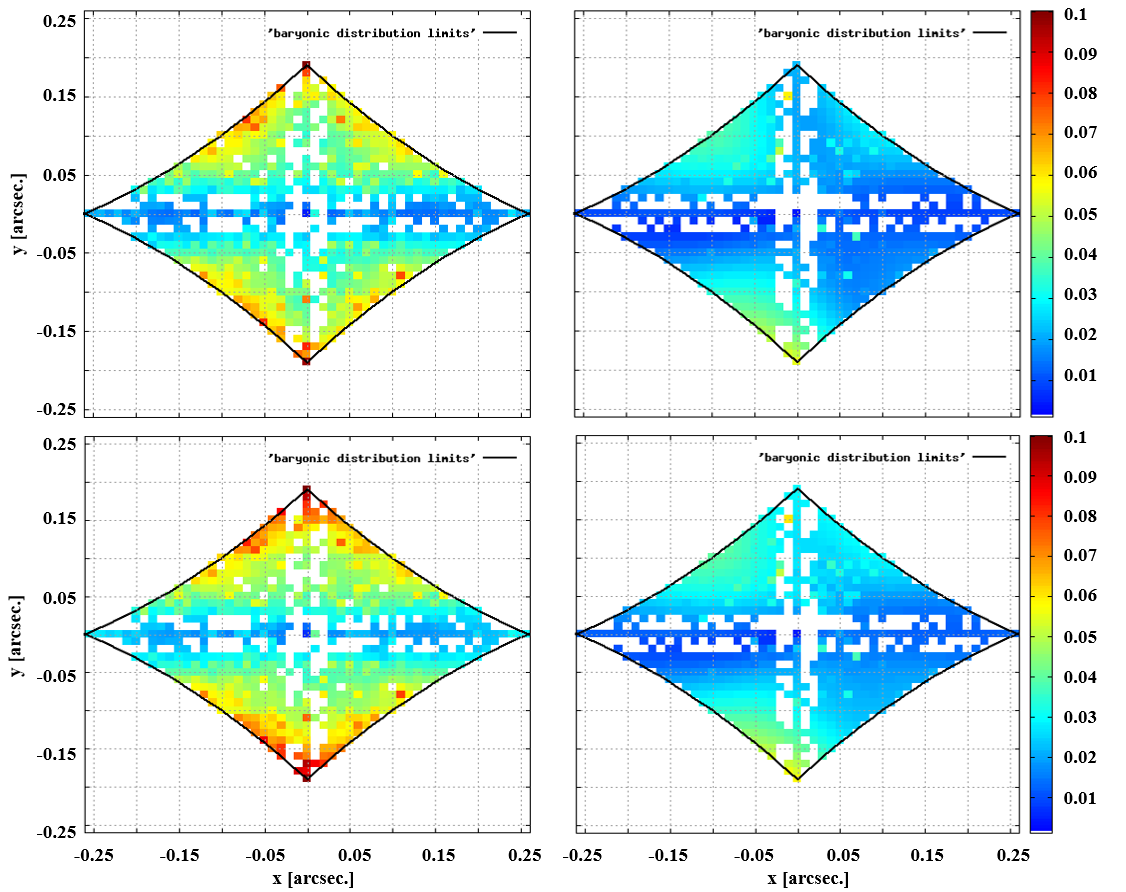}
   	\caption{Unweighted and weighted average $\rm {RMS}$ plots (top and bottom, respectively) for Galaxy 1 (left) and Galaxy 2 (right). The color of each pixel (for a total of 871 per plot) represents the average $\rm {RMS}$ of all points within the extended source. The axes show the offset between the centres of the dark matter and stellar distributions. The black curves demarcate the regions where quad production is guaranteed. The shape of this domain is related to the shape of a diamond caustic, but note that the plane shown is not the source plane. The extended source has a radius $0.04 \arcsec$, and is centered on $(x,y)=(0.0025\arcsec,0.0025\arcsec)$ for Galaxy 1 and $(x,y)=(0.015\arcsec, 0.015\arcsec)$ for Galaxy 2. The $\rm {RMS}$ colour scale is linear, and different from that in Fig.~\ref{fig:plot_src}. The white pixels are instances where \gl~could not produce a 1-component model (cf. Step 3 in Sec. ~\ref{sec:methods}). See Appendix ~\ref{Appendix} for further discussion.}
	\label{fig:avg_plot}
\end{figure*}

\subsection{Extended, $r_s=0.04\arcsec$ sources; offsets $\leq 0.26\arcsec$} 

Each of the eight panels in Fig.~\ref{fig:plot_src} displays the $\rm {RMS}$ values of the 251 individual point-sources (black dots) making up one extended source. The colour scales are the same in all panels, with darkest blue (red) representing $\rm {RMS}$$=0.001\arcsec$ ($0.1\arcsec$).  The four panels on the left (right) show Galaxy 1 (Galaxy 2) lenses, with different dark matter-stellar distribution offsets (see figure caption). These four offsets result in the two lowest, a typical, and one of the highest $\rm {RMS}$ values of the 871 in our analysis. 

The main factor determining the typical $\rm {RMS}$ in these panels is the proximity of the source to the fold or cusp of the diamond caustic: the closer the source to the caustic, the larger the magnification, which means that the point-images comprising the extended image cover more of the lens plane, requiring better model fits for the same lens plane $\rm {RMS}$. Sources closest to caustic centre have lower magnifications, and so are easiest to fit, resulting in lowest $\rm {RMS}$.

The distribution of $\rm {RMS}$ within each extended source (each panel of Fig.~\ref{fig:plot_src}) is less intuitive. Let us focus on one of the eight panels, namely, the top right of the first set of four. The centre of the stellar distribution is at to the right of dark matter centre, so the mass distribution in the lens plane is similar (though not the same) as in Fig.~\ref{fig:massgrid}. When the source is on the left side, three images are on the same side, where the mass is dominated by the dark matter, making it easier to model with a 1-component lens. This is the reason why the left side that panel of Fig.~\ref{fig:plot_src} has low $\rm {RMS}$ values. When the source is one the right side, three images are also on the right, where the mass distribution has competing and uneven contributions from the dark matter and stars, making it hard to model the galaxy with a 1-component lens, resulting in the right side of the panel in Fig.~\ref{fig:plot_src} having higher $\rm {RMS}$ values.

Next, we calculate $\rm {RMS}$ averaged over the extended sources. We used two different schemes to weight the point-sources within an extended source. The `unweighted' scheme weights each of the point-sources (black dots in Fig.~\ref{fig:plot_src}) equally, with the exception of the central point-source, which has to have a weight of 25, one per spoke. So the sum is over the 275 point-sources of the extended source:
\begin{equation}
{\rm RMS}_{\rm ~unwtd} = \frac{1}{275}\,\,{ \displaystyle \sum_{i=1}^{25} \sum_{j=0}^{10} {\rm RMS}_{i,j}}
\label{eq:unwtdavgRMS}
\end{equation}
where ${\rm RMS}_{i,j}$, for one point-source, is given by Eq.~\ref{eq:rms}. This weighting is appropriate for a source whose surface brightness decreases linearly with increasing distance from center. Assuming that brighter pixels would contribute proportionately more constraining power justifies their increased weight. Ideally, the weighting has to also depend on the magnification of each portion of an extended image, but we ignore that here.

The `weighted' scheme is appropriate for a source with a flat surface brightness distibution:
\begin{equation}
{\rm RMS}_{\rm ~wtd} = \frac{ \displaystyle \sum_{i=1}^{25} \sum_{j=1}^{10} r_{j} \times {\rm RMS}_{i,j}}{\displaystyle \sum_{i=1}^{25} \sum_{j=1}^{10} r_{j}}.
\label{eq:wtdavgRMS}
\end{equation}
Here, the central point-sources are not included, and all the pixels within the source have the same weight, regardless of their distance from center. The difference between the two weighting schemes provides approximate uncertainties on the $\rm {RMS}$ values arising from different brightness distributions of the source.

The results are presented in Fig.~\ref{fig:avg_plot}; the two left (right) panels are for Galaxy 1 (Galaxy 2). The two top (bottom) panels show the unweighted (weighted) average $\rm {RMS}$. Each colour pixel represents one extended source, such as the ones shown in Fig.~\ref{fig:plot_src}. The axes show the $x$ and $y$ offset of the centre of the stellar distribution with respect to the centre of dark matter distribution, in the 2-component lens. For example, the four caustics shown in Fig.~\ref{fig:caustics} correspond to the four corners (top-most, bottom-most, left-most and right-most) of the colour rectangles shown here. The black outline that resembles a diamond caustic shows the extreme positions of allowed stellar distribution offsets. For stellar distributions centred outside of these boundaries, the extended source does not fit within the diamond caustic. These cases were not considered. 

A fraction of pixels in the four panels of Fig.~\ref{fig:avg_plot} are white. For these stellar distribution offsets \gl~could not produce a 1-component model, for any of the point-sources in the extended source. These pixels are concentrated along the principal axes, but other than that there is no discernible pattern. Because the overall colour trends are clear despite the white pixels, we will ignore them. (A discussion of what could have gone wrong can be found in Appendix ~\ref{Appendix}.) 

The most prominent feature of these plots is that Galaxy 2 quads are systematically better modeled with 1-component lenses that Galaxy 1 quads. This is because the density profile slope of the 1-component model was fixed at ``isothermal'', $d\ln\kappa/d\ln r=-1$ ($\alpha=1$),  which is  closer to the average slope of Galaxy 2 ($d\ln\kappa/d\ln r\!\sim\! -1.2$, for zero offset) at the image radius, than to that of Galaxy 1 ($d\ln\kappa/d\ln r\!\sim\! -1.25$, for zero offset); see Fig.~\ref{fig:massgrid}.  If the 1-component density slope was allowed to vary during fitting, lower $\rm {RMS}$ values could have been achieved for both Galaxy 1 and 2 quads.

Overall, we see that the lowest average $\rm {RMS}$ values in Fig.~\ref{fig:avg_plot} are near the centres of each of the four panels (as already discussed earlier in this section), which correspond to the smallest dark matter-stars offsets in a 2-component lens. In these cases, the azimuthal mass asymmetry near the image circle is well described by pure ellipticity, and is easily captured by a 1-component {\tt boxypot}. As the offset gets larger, the mass asymmetry near the images deviates more and more from ellipticity, and it becomes harder for a 1-component lens to reproduce the images. The largest $\rm {RMS}$ are found near the very top and bottom of the panels, which correspond to largest vertical offsets. This is because the ellipticity position angle of both the components are such that the isodensity contours are stretched in the horizontal directions, so offsetting the stellar distribution vertically introduces largest density perturbations to the galaxy. 

\begin{figure}
	\includegraphics[width=\columnwidth]{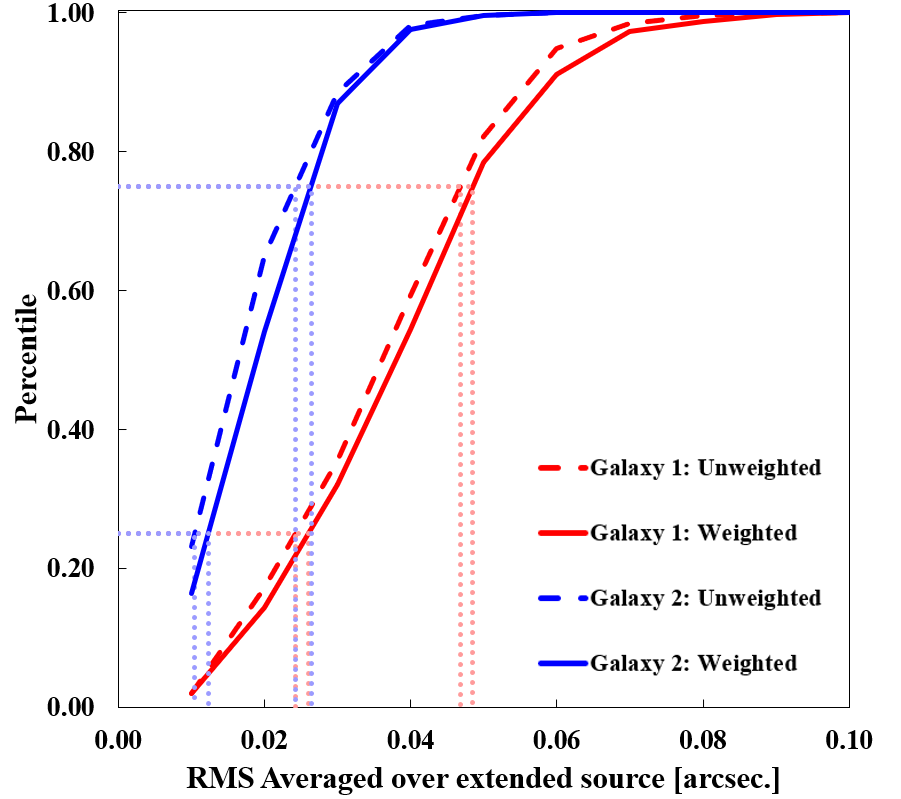}
    \caption{Cumulative distribution functions (CDFs) for Galaxy 1 and Galaxy 2 (red and blue, respectively), and for the unweighted and weighted averaging schemes (dashed and solid lines, respectively). The dotted lines represent the 25th and 75th percentiles.}
    \label{fig:cdf}
\end{figure}

Fig.~\ref{fig:cdf} summarises the $\rm {RMS}$ values of Fig.~\ref{fig:avg_plot} as cumulative probability distributions. In Galaxy 1, $25\%$ ($75\%$) of the extended sources have average $\rm {RMS}$ less than $\sim 0.025\arcsec$ ($\sim 0.05\arcsec$), and so approximately half the sources have the $\rm {RMS}$ values below our threshold of $0.03\arcsec$, for being able to differentiate between a complex 2-component galaxy mass distribution and a simpler 1-component model. In other words, in approximately half of the galaxies with azimuthal mass asymmetries, these asymmetries will go undetected. 
Because we do not consider observational selection biases based on image magnification, these values are to be treated as approximate. 

\subsection{Extended, $r_s\leq 0.14\arcsec$ sources; offsets $=0\arcsec$}

In the previous section we saw that when quads of 2-component galaxies with large offsets (up to $0.26\arcsec$, or $\sim 1h_{0.7}$ kpc) between the centres of their dark matter and stellar distributions are modeled with 1-component lenses, the lens plane $\rm {RMS}$ can reach $\sim 0.07\arcsec$, signaling that simple models do not fully capture the complexity of such galaxies. So for large offsets, source sizes larger than what we assumed there are not required to rule out 1-component models.

However, for smaller offsets, the $\rm {RMS}$ were $\leq 0.03\arcsec$, especially for Galaxy 2. Therefore, for small offsets, it is interesting compute $\rm {RMS}$ values for larger source sizes. We now consider 2-component galaxies with zero offsets, and source sizes that are larger, but such that they do not touch the diamond caustic. The density profile slopes the two 2-component galaxies used here are exactly those shown in Fig.~\ref{fig:slopes}. We note that zero, or near zero offsets is what is commonly assumed in modeling.

\begin{figure*}
	\includegraphics[width=\textwidth]{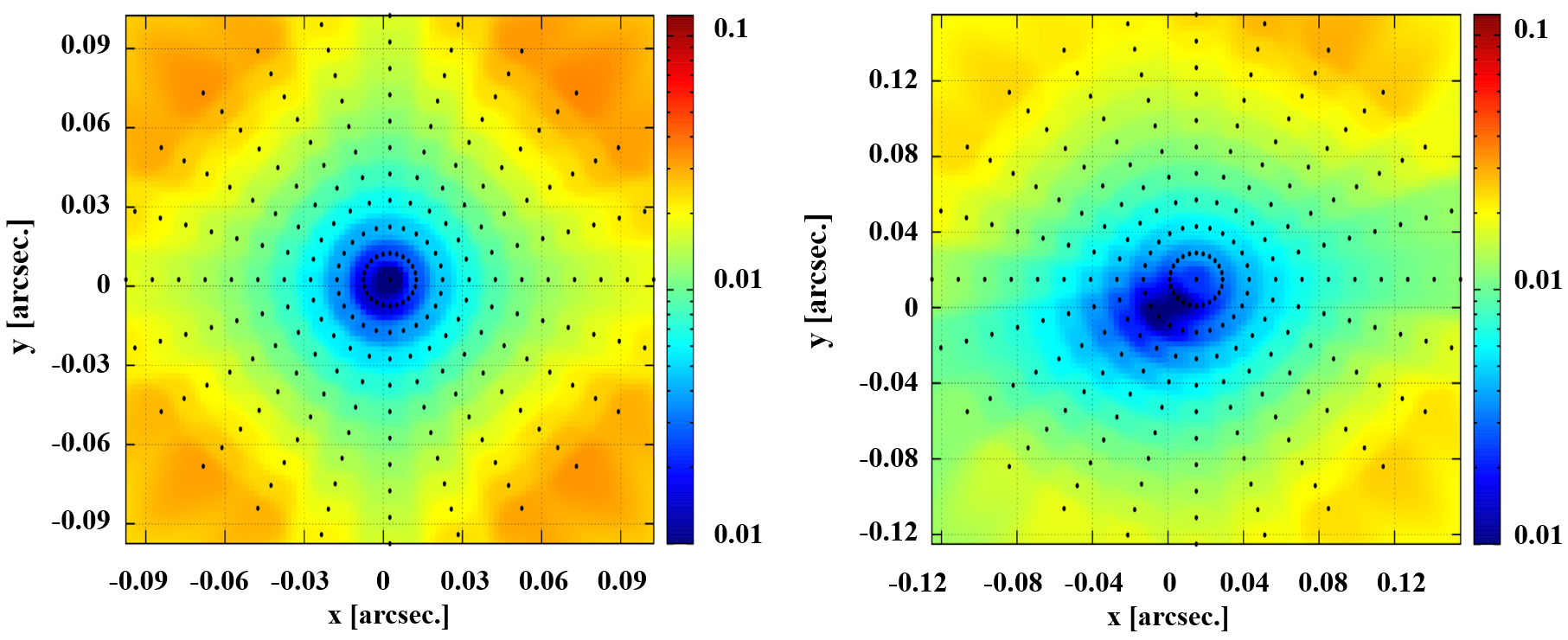}
   	\caption{Similar to Fig.~\ref{fig:plot_src}, but for large sources and 2-component galaxies ({\it left:} Galaxy 1, {\it right:} Galaxy 2) with zero offset between dark matter and stars. The galaxy parameters are given in Table ~\ref{tab:lens_comp} }
	\label{fig:LargeExtExp}
\end{figure*}

The sources used in the previous subsection have radii of $r_s=0.04\arcsec$. Here we use $r_s=0.1\arcsec$ ($\sim\! 0.75h_{0.7}$~kpc) for Galaxy 1 and $r_s=0.14\arcsec$ ($\sim\! 1.05h_{0.7}$~kpc) for Galaxy 2 extended sources and perform the analysis described in Section~\ref{sec:methods}. The sources were placed, as before, at $(x,y)=(0.0025\arcsec, 0.0025\arcsec)$ for Galaxy 1, and at $(x,y)=(0.015\arcsec, 0.015\arcsec)$ for Galaxy 2. The results are presented in Fig.~\ref{fig:LargeExtExp}, for Galaxy 1 (left panel) and Galaxy 2 (right panel). 

\begin{figure}
	\includegraphics[width=\columnwidth]{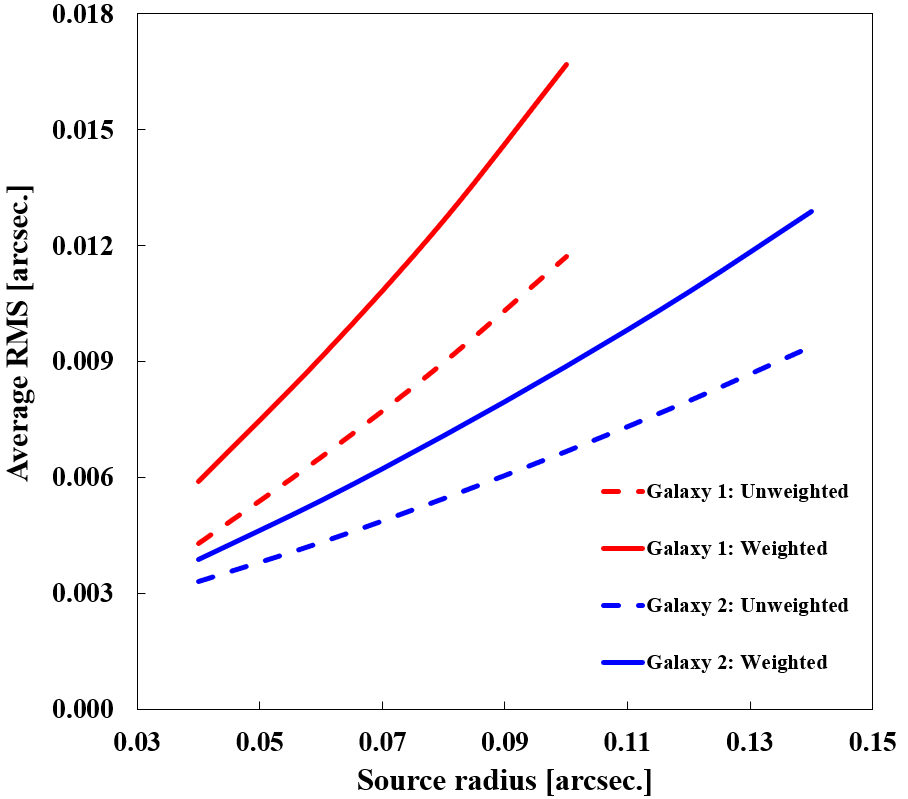}
   	\caption{The source-averaged $\rm {RMS}$, eq.~\ref{eq:unwtdavgRMS} (dashed), and \ref{eq:wtdavgRMS} (solid) as a function of source size, for 2-component galaxies (Galaxy 1 [red], and Galaxy 2 [blue]) with zero offset between dark matter and stars.}
	\label{fig:ExtSrcsExp}
\end{figure}

The outer portions of the sources are harder to model because they are closer to the caustic, but even there, the typical $\rm {RMS}$ values are around $0.03\arcsec$, and so at or below our threshold. The results of these experiments are summarized in Fig.~\ref{fig:ExtSrcsExp}, which shows the average $\rm {RMS}$ as a function of source size, using our two weighting schemes, eq.~\ref{eq:unwtdavgRMS} (dashed), and \ref{eq:wtdavgRMS} (solid). The average $\rm {RMS}$ is seen to increase approximately linearly as a function of source radius. Of the four cases we present here, the largest average $\rm {RMS}$ correspond to the weighted $\rm {RMS}$ case, and the parameters of Galaxy 1. The approximate equation describing the red solid line in Fig.~\ref{fig:ExtSrcsExp} is, $RMS=0.183\times r_s-0.0013$ (in arcsec.) So for zero offset, and sources that fill almost the entire diamond caustic, $r_s\sim 0.15\arcsec$, the resulting images can be well fit with an elliptical 1-component lens, hiding true 2-component nature of the galaxy lens. 

\section{Summary and DISCUSSION}

Even though most individual quads are well represented by simple elliptical+shear lens models, the population of quads is not well represented by such lenses. \cite{GomerWilliams2017} studied the distribution of quads in the 3D space of relative image polar angles, and concluded that fold- and cusp-type quads arise, as expected, from lenses with a range of external shear values, whereas cross-type quads seem to arise only from lenses with small shear.  Cross quads with large shears appear to be missing.  This odd finding, which cannot be accounted for by selection effects, prompted the authors to consider other possible sources of mass perturbation, or asymmetry in the lenses. Having first ruled out $\Lambda$CDM substructures as the reason, they turned to mass asymmetries associated with the transition region, where the central stellar distribution gives way to dark matter. Because the central regions are not fully relaxed, azimuthal asymmetries are expected around this radius. The authors showed that these would give rise to 3D distribution of quad relative image angles consistent with the currently available sample of quad galaxy lenses.

One might then ask, if such mass perturbations exist, why have they not been detected already, through mass modeling? In this paper our goal was to investigate how often, and under what circumstances, these asymmetries can evade detection, based on mass distributions recovered by modeling. 

In our preliminary tests presented in this paper we generated quad images of extended sources, using 2-component---stars and dark matter---galaxies, and then attempted to fit these using 1-component elliptical+shear models. The easiest way to mimic mass asymmetry at the image radius is to offset the two components.  We represented extended sources with dense clusters of point-sources, and compared the ``observed'' images with modeled images, using $\rm {RMS}$ between the corresponding point-images comprising the extended images. Our $\rm {RMS}$ threshold to separate good from bad fits was set at $0.03\arcsec$, somewhat smaller than the HST ACS pixel, because point-images in the lens plane with smaller separations would not be distinguished.


For 2-component galaxy lenses with small or zero offsets, 1-component models can successfully reproduce the images of even the large, caustic-filling sources; i.e., $\rm {RMS}$ were in general below $0.03\arcsec$. In these cases, the fact that a lens has two components will not be revealed by mass modeling. 

For 2-component lenses with non-zero offsets (up to $\sim 1h_{0.7}$~kpc) the source averaged $\rm {RMS}$ depends on several parameters. Sources with flat brightness distributions and further from the caustic centre are worse fit with 1-component models, and hence more likely to indicate the presence of mass perturbations beyond simple ellipticity.  Also, if the average density profile slope of the 2-component galaxy at the location of the images differs from the assumed slope of the 1-component model, such models would have larger $\rm {RMS}$, and also signal the presence of additional mass perturbations. However, there is no guarantee that an improved mass model, one that has a lower $\rm {RMS}$, will recover the true mass distribution adequately well. 

Overall, we estimate that of order half of the quads will be able to eliminate simple models, though not necessarily lead to the true mass distributions. This is not an exhaustive study; a more detailed study would consider a wider range of possible galaxy models, and incorporate the effects of magnification bias.  


In this work we did not use any image time delay information, even though it is available for a subset of lenses, where the central quasar shows variability \citep[e.g.,][]{bon18,slu12}. If all three relative time delays between four images of a quad are known precisely, to better than a few percent, the ratios between these can be used as additional constraints to break degeneracies discussed in this paper, and help isolate the true mass solution. The same quad can then also be used to estimate $H_0$.

However, most of the currently available quads have only one precisely measured time delay \citep[e.g.,][]{cor18}. In such a case, one can use the time delay either to constrain mass model asymmetries, or to estimate $H_0$, not both. Until larger samples of quads with multiple precise time delays become available, the best strategy for obtaining $H_0$ is to use ensembles of lenses and constrain them to share the same $H_0$ value, as was done in \cite{col08,par10,lub12}.

In the present paper, we fitted the central point images of quads, and asked if these fitted mass models can also reproduce extended images within observational uncertainty. For the purposes of $H_0$ estimation, it is now common to use the entire area of extended images to do the mass fitting \citep[e.g.,][]{suy17}, which would require an approach different from the one we implemented here. For all these reasons, $H_0$-related investigation is postponed to a future work. 

\section*{ACKNOWLEDGMENTS}
We would like to thank the University of Minnesota Undergraduate Research Opportunities Program whose support sowed the seeds of this project, allowing it to naturally build, evolve, and come to fruition. We also thank Chuck Keeton for making his versatile and well-documented lens modeling software publicly available.

\bibliographystyle{mnras}
\bibliography{MNRAS31} 


\appendix

\section{ANALYZING FAILED SINGLE-COMPONENT LENSES}	\label{Appendix}
Here we discuss some technical aspects of Fig.~\ref{fig:avg_plot}. 
For most dark matter-stars offset values, \gl~worked as it was supposed to, and generated quads (green crosses in panel {\framebox{\bf 3}} of Fig.~\ref{fig:algorithm}),  starting from the average source position (blue triangle), and using the best-fitting 1-component lens (violet oval). 

However, for some offset values, which we represented by white pixels in Fig.~\ref{fig:avg_plot}, \gl~did not produce quads, despite the fact that our original extended source was lensed successfully by \gl~through the 2-component galaxy (panel {\framebox{\bf 1}} of that figure).  This is surprising because the differences in most parameters vary smoothly from one offset value (one pixel in Fig.~\ref{fig:avg_plot}) to the next. 

We will examine two adjacent offset values,  $(x,y) =(0\arcsec,0\arcsec)$ and $(-0.01\arcsec,0\arcsec)$. 
One reason \gl~might be able to produce images with a 2-component, but not a 1-component lens, would be that the back-projection process incorrectly yielded an extended source that was wildly different from the original. We checked; this was not the case. 

We speculate that the two most likely culprits are either the $x$ and $y$ positions of the stellar distribution (i.e., the offset), or the lens parameter values. We tested these two possibilities, in turn.

The parameters of the two 1-component models, with two adjacent offsets, are presented in Table ~\ref{tab:single_comp_comparison}. The values of $b$ and $\epsilon$ for both lenses are very close  to each other, and the values of $\alpha$ and $\beta$ are exactly the same, so none of these parameters are likely to be the problem.  The two values of $\theta_{\epsilon}$ are different by nearly $\sim{180}^\circ$, which means that the orientation of the ellipticity position angle is nearly the same in both cases.  

\begin{table}
	\centering
	\caption{Example 1-component {\tt boxypot} lenses (found by \lm) for Galaxy 1. The top row, with the stellar distribution centred at $(x,y)=(0.00\arcsec, 0.00\arcsec)$, resulted in the successful production of a quad. The bottom row, with stars centred at $(x,y)=(-0.01\arcsec,0.00\arcsec)$, produced no images. Values for $\alpha$ and $\beta$ were kept constant at $1.0$ and $0.5$, respectively.}
	\label{tab:single_comp_comparison}
	\begin{tabular}{ccccccc} 
		\hline
		 x [\arcsec] & y [\arcsec] & $b$ & $\epsilon$ & $\theta_{\epsilon}[^\circ]$ & $\gamma$ & $\theta_{\gamma} [^\circ]$  \\
		\hline
		 0.00 & 0.00 & 1.80 & 0.0540 & -89.9 & 2.63e-04 & 86.1\\
		 -0.01 & 0.00 & 1.81 & 0.0569 & 90.0 & 5.45e-05 & -32.1\\
		\hline
	\end{tabular}
\end{table}


Since we know the 1-component models with the stars centred at $(x,y)=(0\arcsec,0\arcsec)$ does produce a quad, we changed its lens parameters to those of the 1-component model with  $(x,y) = (-0.01\arcsec,0\arcsec)$ offset, and attempted to produce a quad. Changing the values of $b$, $\epsilon$, $\theta_{\epsilon}$,$\gamma$, and, $\theta_{\gamma}$ individually, and collectively resulted in a quads being produced in every instance. So lens model parameters are not the problem. 

We then changed the offset of the 1-component lens originally at $(0\arcsec,0\arcsec)$ to $(-0.01\arcsec,0\arcsec)$. That failed to produce quads.  This indicates that the inability of \gl~to produce quads is related to the value of the offset used.
To further test that hypothesis, we attempted to produce a quad assuming all lens parameters associated with the $(-0.01\arcsec,0\arcsec)$ offset, but changing the offset to $(-0.0025\arcsec,0\arcsec)$. This did produce a quad. 

While we now know that varying the location of the stellar distribution seems to be a limitation of \gl/\lm, resulting in failed white pixels in Fig.~\ref{fig:avg_plot}, we are unsure of the reason behind it. The \gl/\lm~manual \citep{gravlens}, while extensive, does not mention the behavior we have observed.


\bsp	
\label{lastpage}
\end{document}